\documentclass[twocolumn,showpacs,english,prl,superscriptaddress,preprintnumbers,amsmath,amssymb,floatfix]{revtex4-1}

\usepackage[T1]{fontenc}
\usepackage[latin9]{inputenc}
\usepackage{dcolumn}
\usepackage{bm}
\usepackage{graphicx}
\usepackage{color}
\usepackage{esint}
\usepackage{babel}
\usepackage{amsfonts}
\usepackage{slashed}
\usepackage{enumerate}

\newcommand{\bk}{{\bf k}}
\newcommand{\bb}{{\bf b}}

\newcommand{\bx}{{\bf x}}
\newcommand{\sn}{\sigma^{\mathrm{n}}}
\newcommand{\boldgreek}[1]{\mbox{\boldmath$\mathbf{#1}$\unboldmath}}
\newcommand{\bsigma}{\boldgreek{\sigma}}
\newcommand{\sla}[1]{\slashed{#1}}

\begin{document}

\title{Effective field theory of the disordered Weyl semimetal}

\author{Alexander Altland and Dmitry Bagrets}
\affiliation{ Institut f\"ur Theoretische Physik, 
Universit\"at zu K\"oln, Z\"ulpicher Str.~77, D-50937 K\"oln, Germany }

\date{\today}

\begin{abstract}

In disordered Weyl semimetals, mechanisms of topological origin lead to the
protection against Anderson localization, and at the same time to different
types of transverse electromagnetic response -- the anomalous Hall, and chiral
magnetic effect. We here apply field theory methods to discuss the
manifestation of these phenomena at length scales which are beyond the scope
of diagrammatic perturbation theory. Specifically we show how an interplay of
symmetry breaking and the chiral anomaly leads to a field theory containing
two types  of topological terms. Generating the unconventional response
coefficients of the system, these terms remain largely unaffected by disorder,
i.e. information on the chirality of the system remains visible
even at large length scales.

\end{abstract}


\maketitle
Weyl semimetals are paradigmatic examples of gapless topological condensed
matter systems. A Weyl semimetal comprises an even number of linearly
dispersive band touching points embedded in a three-dimensional Brillouin
zone. The presence of these hot-spots implies a response to perturbations that
is intermediate between that of metals and insulators (see
Ref.~\cite{Hosur2013Recent} for review). This
`semi-metallicness' also signifies in the physics of the disordered
system~\cite{Sbierski2014Quantum,Burkov2014Chiral,Syzranov2014Critical,Syzranov2014Localisation}: on the one hand, the vanishing of the nodal density of states weakens disorder scattering cross sections; on the other hand, sufficiently
strong disorder will \emph{generate} a finite band center  density of states to
eventually overpower the above effect. It has been
shown~\cite{Syzranov2014Critical} that the above
competition manifests itself in the presence of a critical disorder strength
below/above the system flows towards a clean fixed point/a regime of strong
impurity scattering.  It is the purpose of the present paper to derive and
discuss the effective  theory describing the latter phase
at length scales exceeding the system's  scattering mean free path.

At large length scales, impurity scattering will render the motion of
individual excitations diffusive which suggests that the system will end up
in the universality class of the 3d Anderson metal (i.e. above the phase
transition point separating a 3d metal from an insulator.) This expectation
is, in fact, a certainty given that a single Weyl node may be interpreted as
an effective surface theory of a bulk 4d topological insulator; finite
conduction is protected by topology. At the same time, topology implies a
number of differences distinguishing the Weyl system from a generic metal:
First, an individual Weyl node breaks parity symmetry, and it is known (e.g.
from the example of the $d$-wave superconducting quasiparticle system~\cite{Altland2002Theories})
that the breaking of discrete symmetries is generally remembered, even in the
presence of strong disorder. Indeed, we will find that the low energy
theory of individual nodes system contains a parity breaking non-abelian
Chern-Simons (CS) term, which describes the survival of the so-called chiral
magnetic effect (CME)~\cite{Fukushima:2008,Chen:2013} in the disordered environment. Second, it has been
shown that a system comprising two Weyl nodes separated in momentum space
shows an anomalous Hall effect (AHE)~\cite{Burkov2011Weyl}. Within the field theoretic framework
below, this effect will derive from a 3d extension of a two-dimensional
topological $\theta$-term, familiar from the theory of the quantum Hall
effect.

\begin{figure}
\centering
\includegraphics[width=5cm]{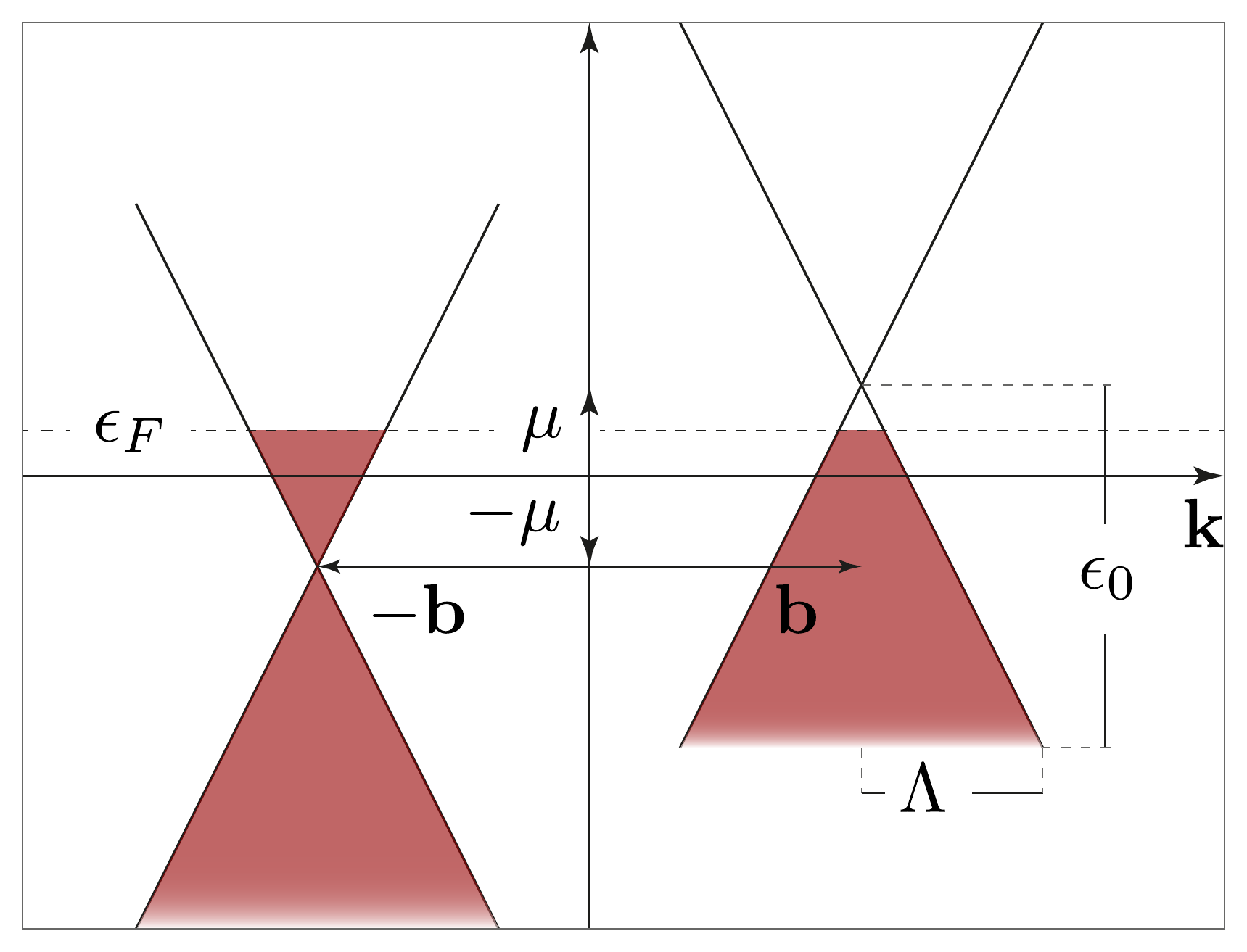}
\caption{\label{WeylNodes} (Color online) 
Schematic of two Weyl nodes split in energy and momentum by $2\mu$ and $2\bb$,
resp.}
\end{figure}

\noindent\emph{Field theory ---}  In the following, we will derive an
effective feld theory describing these
structures. Our starting point is 
the bi-nodal Hamiltonian (cf. Fig.~\ref{WeylNodes})
\begin{align}
 	\hat H = v \sla{\hat{k}}\, \sn_3 +(v\sla{b} + \mu)+  V
(\bx),
 \end{align} 
 where $\sla {\hat k}\equiv \bk\cdot
\bsigma$,  $\bsigma$ is a vector of Pauli matrices, $\hat \bk$ the vector
momentum operator, and $v$ a characteristic velocity. The Pauli matrix $\sn_3$
acts in a two-component space discriminating between two nodes split by a
vector $2\bb\equiv 2b\mathbf{e}_3$ in momentum space and an  increment $2\mu$
in energy. The model is coupled to disorder by a  Gaussian distributed
potential $V (\bx)$ with variance $\gamma_0$. We discriminate between disorder
correlated over length scales $\gtrsim b^{-1}$, which is soft in the sense
that the two Weyl nodes are not coupled by impurity scattering, and the
opposite case of short range correlated disorder mixing the nodes. Implicit to
the model is a high momentum cutoff $|\bk|<\Lambda$ limiting the range of
linearizability of an underlying lattice model.

To access the transport properties of the system at energies $\epsilon$, we
introduce a supersymmetric generating functional $Z[a]=\int D(\bar
\psi,\psi)\,\exp(-S[\bar \psi,\psi])$, with action
$
	S[\bar \psi,\psi,a]=-i\int d^3x \,\bar \psi(\epsilon+i \delta \tau_3 - \hat H)\psi.
$
Here $\psi=\{\psi_{s,i,n}^\alpha(\bx)\}$ is a sixteen-component field of
integration variables where, $n=1,2$ labels the two nodes,  $i=1,2$ the
components of a Weyl spinor, $s=\pm$ distinguishes between advanced and
retarded (ar) Green functions, i.e. $ (\tau_3)_{ss'}=s \delta_{ss'}$ such that
integration over $\psi_{s=\pm}$ generates matrix elements of $\hat
G^\pm=(\epsilon\pm i\delta -\hat H)^{-1}$, and $\alpha=1,2$ defines a
two-dimensional space (bf-space) of complex commuting/Grassmann variables
$\psi^{1/2}$. In the commuting sector, $\bar\psi^1=\psi^{1\dagger}$, while the
commuting variables $\bar \psi^2,\psi^2$ are independent Grassmann variables.
From $Z$, transport observables may be computed by introducing
suitably defined source fields $a$, which are presently suppressed to keep the
notation
simple.

 To
explore the influence of disorder on the system, we integrate over $V$ to
generate the quartic  contribution $\frac{\gamma}{2}\int dx\,
(\bar\psi\psi)^2.$ The fate of this nonlinearity under changes of the
cutoff $\Lambda$ has  been studied~\cite{Syzranov2014Localisation,Roy:2014}  by evaluating the results in one- and
two-loop renormalized perturbation theory in $2+\epsilon$
dimensions~\cite{Fradkin1986Critical}
at $\epsilon=1$. It has been found that for bare amplitudes larger
than a
critical value $\gamma^\ast= \pi^2 v^2/\Lambda$ the effective disorder
strength increases under renormalization. To describe the perturbatively
inaccessible regime beyond the scattering mean free path, $\Lambda^{-1}\equiv
l\sim \gamma/v^2$,  we decouple the nonlinearity by a supermatrix field $B$, and
integrate over the then quadratic fermions~\cite{Efetov1997Sypersymmetry}. As
a result, we obtain the  effective action $S[B]=\frac{1}{2\gamma}\int
d^3x\,\mathrm{str}B^2-\mathrm{str}\ln(\hat G[B])$, where $\hat G[B]=(\epsilon + i \delta
\tau_3 -\hat H_0-  B)^{-1}$, $\hat H_0$ is the clean Hamiltonian and
$\mathrm{str}$ is the supertrace~\cite{Efetov1997Sypersymmetry}. The
difference between the cases of hard and soft disorder, resp., is that in the
former/latter case the two nodes couple to the same ($B=B \otimes
\Bbb{I}^\mathrm{n}$)/independent ($B=\mathrm{bdiag} (B_1,B_2)^\mathrm{n}$)
matrix fields. For definiteness, we first consider the soft case, for which the
two nodes can be discussed separately;
the effect of impurity mixing can be described by a locking $B_1=B_2$ at any
later stage. Focusing on  node $n=1$ and  writing $B=B_1$ for
notational simplicity, we proceed by subjecting the action to a mean field
analysis. A variation of the action yields the equation $\bar
B\stackrel{!}{=}\gamma\,{\rm tr}_{\rm }\,\hat G({\bf x,x};[\bar B])$, which is
solved~\cite{Fradkin1986Critical} by
the diagonal ansatz, $\bar B=-i \kappa
\tau_3$ describing the `spontaneous symmetry breaking' of the infinitesimal
$\delta$ a finite impurity self energy $\kappa=\kappa(\epsilon)$.
Specifically, at
$\epsilon=0$
(semimetal) one
finds
$\kappa = (2/\pi)
v\Lambda (1-{\gamma^\ast}/{\gamma})$~\footnote{For
$\gamma<\gamma^\ast$, the mean field equations have only a trivial solution
$\kappa=0$, which is consistent with the presence of a critical point at $\sim
\gamma^\ast$.}, while far away from Weyl node (metal) $\kappa =
\gamma\pi
\nu$, where
$\nu = \epsilon^2/2\pi^2 v^3$ is the clean density of states.

Turning to the analysis of fluctuations around the mean field,
we focus on configurations $B\equiv i\kappa Q\equiv i\kappa T\tau_3 T^{-1}$,
whose action vanishes in the limit of slowly varying fluctuation matrices
$T(\bx)\to\mathrm{const.}$ Here, $T\in G/H$ takes values in a Goldstone mode
coset space where $G=\mathrm{GL} (2|2)$ is the group of $4\times 4 $
invertible supermatrices and $H=\mathrm{GL} (2|2)$ the subgroup of matrices
$k$ commutative with the mean field, $ [k,\tau_3]=0.$ Since
$\mathrm{str}(Q^2)=0$, the
action we now need to consider reads $S_0[Q]\equiv \mathrm{str}\ln(\epsilon-v\sla k +
i \kappa Q)$. For later reference, we note that the action has two continuous
symmetries: a global symmetry $Q\to T_0 Q T_0^{-1}$, where
$T_0\in G$ is constant, and a local gauge symmetry $T\to Tk$, $k=k(\bx)\in
H$.

\noindent\emph{Effective field theory ---} Our goal is to expand the action in
gradients $T^{-1}\partial_i
T\equiv A_i$.
One might be tempted to start this program with a  similarity transformation,
$S_0[Q]\stackrel{?}{=}\mathrm{str}\ln(T^{-1}(\epsilon-v\sla k + i \kappa
Q)T)=\mathrm{str}\ln(\epsilon-v\sla k+ i v \sla A + i \kappa \tau_3)\equiv S[A]$,
which may then be expanded in powers of the `non-ablian gauge field' $A$.
However, due to the notorious anomaly, this operation is invalid, the action
needs to be  regularized first. Following a strategy previously applied to the
2d $d$-wave superconductor, we regularize by subtraction of a term $S[Q]\equiv
S_0[Q]-S_\eta[Q]$, where $S_\eta$ differs from $S_0$ by a replacement
$\kappa\to \eta\searrow 0$ and setting $\epsilon = 0$. In the limit $\eta\to 0$ the $Q$-dependence of
$S_\eta[Q]$ drops out so that $S_{\eta\to 0}[Q]=0$. On the other hand, for
large momenta $v|k|\gg q$ and fixed $\eta$, the two action contributions
cancel against each other, i.e. $S[Q]$ is UV regularized (in a gauge-invariant
way.) The similarity transformation may now safely be applied to both
$S_{0,\eta}$, so that we obtain an effective action
$S[Q]=(S_0[A]-S_\eta[A])_\mathrm{reg}$, where the subscript `reg' means that
only UV finite contributions to the subsequent expansion in $A$ are to be
kept. Notice that in the language of the $A$-fields, the local gauge
invariance is no longer manifest. Rather, $A_i\to k^{-1}(A+ k \partial_i
k^{-1})k$ transforms as a non abelian gauge field, and invariance of the
action becomes a non-trivial consistency condition.

In the expansion of the action, we keep terms of order two ($\mathcal{O}
(\partial A,A^2)$) and three ($\mathcal{O}(A^3,\partial A)$) derivatives. To
second order we obtain the result
\begin{align}
	\label{S3dsigma}
	S_\mathrm{d}[A]&=-\frac{\sigma^1_{xx}}{8}\sum_i \int d^3 x\,\mathrm{str}(
	[A_i,\tau_3]^2),\crcr
	S_\mathrm{top}[A]&=\frac{\sigma^1_{xy}}{2}\epsilon^{3ij}\int d^3x\,
	\mathrm{str}(\tau_3 \partial_i A_j),
\end{align}
where the longitudinal and Hall conductivity of node 1, $\sigma^1_{xx}$ and
$\sigma_{xy}^1$ are determined by the microscopic model parameters as
discussed below. We note that the action~\eqref{S3dsigma} affords the
manifestly gauge invariant reformulation $S_\mathrm{d} [Q]=
-\frac{\sigma_{xx}}{8}\int d^3 x\,\mathrm{str}(\partial Q^2)$ and
$S_\mathrm{top}[Q]=\frac{\sigma_{xy}}{8}
 \epsilon^{3ij}\int dx\,\mathrm{str}
(Q\partial_i Q \partial_j Q)$. The first of these contributions has been
constructed in Ref.~\cite{Fradkin1986Critical} on phenomenological grounds
within a non-regularized framework. (In our current approach regularization
plays a vital role; the `cross term' $A\tau_3A\tau_3$ of the commutator
$[A,\tau_3]^2$ is weighed with an UV divergent `fermion bubble'. Only after
regularization it combines with the finite coefficient of $A^2$ to a gauge
invariant commutator.) Before discussing the physics of these expressions, we
complete the derivation of the action and consider terms of cubic order in
$A$.

The terms of $\mathcal{O}(A^3)$ are the `triangle graphs' pervasive in the theory
of $(2+1)$ or $(3+0)$ dimensional relativistic gauge theories. On general
grounds~\cite{Redlich1984Parity} we expect the appearance of a Chern-Simons 
action
at this order. A straightforward if lengthy calculation  indeed yields
the result
\begin{align}
	\label{SCS}
	&S_{\mathrm{CS}}[A]=S^I_{\mathrm{CS}}[A]+S^{II}_{\mathrm{CS}}[A],\\
	&\;S^I_{\mathrm{CS}}[A]=\frac{i\epsilon^{ijk}}{8\pi}\sum_{s=\pm} s\int
	d^3x\,\mathrm{str}
	(A_i P^s
	\partial_j A_k P^s),\crcr
	&\;S^{II}_{\mathrm{CS}}[A]=\frac{i\epsilon^{ijk}}{12\pi}\sum_{s=\pm} s\int
	d^3x\,\mathrm{str}
	(A_i P^s
	A_j P^s A_kP^s),\nonumber
\end{align}
where $P^\pm$ is a projector on advanced/retarded indices.
Apart from the presence of these projector matrices, this has the characteristic
structure of a non-abelian CS term. (However,~\eqref{SCS} does not
define a `real' CS action, inasmuch as $A$ does not describe a field-gauge
coupling, but represents the nonlinear $\sigma$-model field itself; the
situation
is conceptually similar to  that
considered in
Refs.~\cite{Wilczek1983Linking,Volovik1989Fractional}.)

The CS action does not afford a representation in terms of $Q$-fields, which
reflects the lack of complete gauge invariance of this action
piece~\cite{Redlich1984Parity}: it is straightforward to verify that under a
gauge transformation by $k\equiv \mathrm{bdiag }(k_+,k_-)^\mathrm{ar}\in H$, and for a fictitious
system without boundaries, the CS action transforms as $S_\mathrm{CS}[A]\to
S_\mathrm{CS}[A]+S_\mathrm{top}[k]$, where $S_\mathrm{top}
[k]=\frac{-i}{24\pi}\sum_{s=\pm} s\int d^3x\,\mathrm{str}
(k_s^{-1}\partial k_s)^{\wedge 3}$. The integral
yields a quantized value, viz. $24\pi^2\times n_s$, where $n_s$ is the is
winding number of a configuration $k_s^\mathrm{ff}\in \mathrm{SU}(2)$ in three
dimensional space. (The non-compact bb sector of the supermatrices $k$ is
topologically empty.) For `large' gauge transformations with non-vanishing
winding numbers, the CS action changes by a factor $i\pi (n_+ + n_-)$. The
origin of this
phenomenon was explained in the classic reference~\cite{Redlich1984Parity},
where it was shown that these factors get canceled by a gauge anomaly of the
regulator action 
$S_\eta\to S_\eta + i \pi(n_++n_-)$, where the appearance of the extra terms
is caused by zero-crossings of the regularizing Dirac operator under a large
gauge transformation. Overall the action is gauge invariant.

\noindent \emph{Discussion ---} 
The action $S_d + S_\mathrm{top}$  of the diffusion modes $A$ appeared for the
first time in connection with the multi-layer quantum Hall
effect~\cite{Wang1997Localization}, a system conceptually similar to the
present one if the $3$-direction of Weyl node splitting is interpreted as the
stacking direction of a layered system of $2d$ class A topological insulator
layers~\cite{Burkov2011Weyl}. The action $S_\mathrm{d}$ controls the fluctuations of diffusion
modes in terms of the dimensionless coupling constant $g_{xx}\equiv
\sigma^1_{xx}
\Lambda^{-1}$
Within the framework of our gradient expansion we find
$\sigma_{xx} = (\epsilon^2 + 3
\kappa^2)/ 6\pi\kappa v$, which simplifies to
$\sigma^1_{xx} = \kappa/2\pi v$ at the Weyl node and asymptotes to the Drude
conductivity $\sigma^1_{xx} = v^2/3\gamma$ at higher energies $\epsilon \gg
\kappa$. At the nodes, $\epsilon=0$, and at bare length scales $\Lambda\sim
l^{-1}$ characteristic for the
ballistic/diffusive crossover, the conductance $g_{xx}$ takes
values of $\mathcal{O}(1)$, close to but larger~\footnote{Evidence that we
start out on the metallic side of the transition follows from (i) the
topological protection against localization, (ii) matching with the RG flow of
the underlying $\psi$-field theory  which is directed towards metallic
behavior, and (iii) the fact that the RG analysis of the $\sigma$-model in
$d=2+\epsilon$, $\epsilon=1$, dimensions predicts a value (cf.
e.g.~Ref.\cite{Wang1997Localization}) ${g^\ast}={1}/{\sqrt 3 \pi^2}$.} than
the critical value $g^\ast$ marking the 3d Anderson transition. 

The bare coefficient $\sigma^1_{xy}$ multiplying the action $S_\mathrm{top}$
is the contribution of node 1 to the Hall conductivity of the system at
crossover length scales to the diffusive regime. As in the derivation of the
field theory of the quantum Hall effect~\cite{Levine1984Theory} we obtain
$\sigma_{xy}$ as a thermodynamic coefficient $\sigma_{xy}=V^{-1}\partial_B N$,
where $V\equiv L_z L^2$ is the volume of the sample, and the
derivative probes
the sensitivity of the number of states
$N=-\frac{1}{\pi}\int_{-\infty}^0d\epsilon\, {\rm Im}\,
\mathrm{tr}(G^+(\epsilon))$ below zero energy
to the presence of an external magnetic field in $3$-direction. Within the
framework of the gradient expansion, $G^+$ is
the retarded Weyl node Green function coupled to disorder in the mean field/self
consistent Born approximation. Following the strategy of
Ref.~\cite{Burkov2011Weyl}, we evaluate this expression by interpreting the
quantized values of the $3$-momentum, $m_n\equiv v(b+k_{3,n})$, $k_3=2\pi
nL_z^{-1}$ as masses entering effective $2d$ layered Dirac Hamiltonians
governing the planes perpendicular to the $3$-direction. Within this
interpretation, $\sigma_{xy}=(2\pi/L_z)\sum_n \sigma_{xy,n}$ is obtained by
summing over the contribution of the layers, where $\sigma_{xy,n}=C_n/2\pi$ is
given by the disorder averaged Chern number, $C_n = (1/2\pi)\sum_{\sigma = \pm }
\arctan\left[(m_n+ \sigma \epsilon)/{\kappa}\right]+1/2$, of the $n$th layer. Since only
\emph{changes} of $C_n$ at zero crossings of the effective masses can be
unambiguously determined from the linearized theory, we fix the absolute value
of the sum by the condition
$\sigma_{xy} = 0$ at $b=0$, which follows from  matching to the 
band structure of the bi-nodal Weyl system.  In the limit $L_z b \gg 1$, this
gives $\sigma_{xy}^{1/2}=b/2\pi$ for both nodes, irrespective of the energy
$\epsilon$ or the disorder strength $\kappa$. Both
for soft and hard disorder, the two contributions to the Hall conductivity add,
and we obtain $\sigma_{xy}=b/\pi$, a result known  as the anomalous Hall effect (AHE).

The CS contribution to the action accounts for the \emph{thermodynamic}
response of
the
system to imbalances ($\bb,\mu$) between the nodes. That this term probes
equilibrium properties follows from the presence of the projector matrices
$P^s$ which
prevent coupling between the retarded and advanced sector of the theory (a
necessary ingredient to any type of dynamic response.) A finite equilibrium
response is obtained if we couple the system to an external field $a=\{a_i\}$,
where $a_i=\frac{B}{2}\epsilon_{3ij}x^j$, $i=1,2$ represents an external
magnetic field $B\mathbf{e}_3$, and $a_3=a(x) \sigma_3^\mathrm{bf}\otimes
\tau_3$ is a
source field.
The latter is defined in such a way that differentiation of the partition
function, $\frac{i}{4\pi}\delta_{a(\bx)} Z
[a]=-\frac{1}{\pi}\mathrm{Im}\,\langle
\mathrm{tr} [G^+
(\bx,\bx)\sigma^3]\rangle =\langle j_3(\bx)\delta(\epsilon-\hat
H)\rangle\equiv  j_{3,\epsilon}$ yields the contribution of
states at energy $\epsilon$ to the equilibrium value of the $3$-current
density of node 1. We compute this expression by adding the external field to
the internal one, $A\to A+a$, and substituting this configuration into the CS
action. In the simplest approximation $A=0$ (for the above `equilibrium'
choice of source terms fluctuation corrections around the $A=0$ vanish by
supersymmetry anyway), we then obtain $S_\mathrm{CS}[a]=-\frac{iB}{\pi}\int
dx\,a(\bx) $, and hence $ j_{3,\epsilon}= \frac{1}{4\pi^2} B$. To obtain the
full response of the system, we need to add
the (opposite) contribution of the second node and integrate over filled
energy states up to some Fermi energy $\epsilon_F$. Taking into account that the
existence of a bare linearization cutoff $\Lambda$ implies a
cutoff $|\epsilon|<|\epsilon_0\pm \mu|$, $\epsilon_0\equiv v \Lambda$ for the
accessible energy states
(cf.
Fig.~\ref{WeylNodes}), this leads to $j_3=\int^{E_F}_{-\epsilon_0+\mu} d\epsilon
j_{3,\epsilon}-\int_{-\epsilon_0-\mu}^{\epsilon_F} d\epsilon\,
j_{3,\epsilon}=\frac{\mu B}{2\pi^2}$, i.e. an equilibrium current proportional
to an external magnetic field, the so-called chiral magnetic effect (CME)~\cite{Fukushima:2008,Chen:2013}
(for a discussion how the nonvanishing of this expression may be understood from the perspective of the Fermi-liquid theory, see Ref.~\cite{Yamamoto:2012}).

\noindent{\emph{Renormalization} --} What happens if short distance fluctuations
in
the field
theory are
integrated
out to probe the physics at length scales beyond the ballistic/diffusive
crossover regime? An answer to this question has been formulated in
Ref.~\cite{Wang1997Localization} within the framework of two loop renormalized
perturbation theory for the dimensionless coupling constants
$g_{\mu\nu}=\sigma_{\mu\nu}\Lambda^{-1}$ of the model. The result
\begin{align}
	\frac{d g_{xx}}{d\ln l}=g_{xx}-\frac{1}{3\pi^4 g_{xx}},\qquad
	\frac{d g_{xy}}{d\ln l}&=g_{xy},
\end{align}
states that the longitudinal conductance scales according to the predictions
of one-parameter scaling theory (unaffected by the Hall conductance) towards
Ohmic behavior $g_{xx}\stackrel{g_{xx}\gg 1}\sim l$. The Hall conductance
shows linear scaling, $g_{xy}\propto l$, which means that the AHE remains
unrenormalized by disorder $\sigma_{xy}=\mathrm{const.}$ even at large length
scales (in contrast to the Hall conductivity of a two-dimensional system which
is renormalized by instanton fluctuations~\cite{Levine1984Theory}). Finally,
the coupling constant of the CS action is fixed by gauge invariance, and
fluctuation corrections to $\langle j_3\rangle_B$ vanish by supersymmetry.
This means that within the framework of our theory the CME, too, fully
protected against renormalization by disorder.

The disorder insensitivity of the topological response coeffecients holds
regardless of whether we are probing the semimetallic Weyl nodes
$\epsilon,\mu\sim \kappa$,
or the metallic physics far away from them $|\epsilon\pm \mu|\gg \kappa$. The
essential difference between the two situations lies in the bare and
renormalized values of the longitudinal conductance $g_{xx}$: in the
former/latter case, $g_{xx}$ is initially small/large to begin with. However, in
either case, $g_{xx}$ increases, and asymptotes to Ohmic behavior at large
length scales. While the system then behaves similar to a three dimensional
metal, the
preserved non-vanishing of its two transverse transport coefficients betrays
the underlying presence of two Dirac nodes.

Summarizing, we have microscopically derived a supersymmetric field theory
description of disordered Weyl semimetals and metals at length scales exceeding
the mean free path. The profile of the theory is essentially determined by an
interplay of symmetry conditions and the chiral anomaly. This mechanism
stabilizes metallic behavior at large length scales, along with various
disorder-insensitive response coefficients of topological origin.

\noindent{\emph{Acknowledgments:} We acknowledge discussions with P. Brouwer,
M. Hermanns, and S. Trebst. Work supported by SFB/TR 12 of the Deutsche
Forschungsgemeinschaft. In the final stages of the preparation of this paper
we became aware of arXiv:1412.7678 where an action similar to ours is motivated
from a different perspective, viz. by dimensional reduction from a bulk
$4d$-topological insulator.}

\bibliography{Bibliography} 

\end{document}